\documentclass[10pt]{iopart}

\usepackage{graphicx}

\newcommand{\be}{\begin{equation}}
\newcommand{\ee}{\end{equation}}
\newcommand{\bea}{\begin{eqnarray}}
\newcommand{\eea}{\end{eqnarray}}
\newcommand{\eps}{\epsilon}

\begin{document}

\title{Applications of distance between probability
distributions to gravitational wave data analysis}
\author{Robert J. Budzy\'nski}
\address{Department of Physics, Warsaw University, Ho\.za 69, 00-681 Warsaw, Poland}
\author{Witold Kondracki and Andrzej Kr\'olak}
\address{\em Institute of Mathematics, Polish Academy of Sciences, \'Sniadeckich 8,
00-956 Warsaw, Poland}

\begin{abstract}
We present a definition of the distance between probability
distributions. Our definition is based on the $L_1$ norm on space of
probability measures. We compare our distance with the well-known Kullback-Leibler
divergence and with the proper distance
defined using the Fisher matrix as a metric on the parameter space. We
consider using our notion of distance in several problems in gravitational
wave data analysis: to place templates in the
parameter space in searches for gravitational-wave signals, to assess quality
of search templates, and to study the signal resolution.
\end{abstract}

\maketitle

\section{Introduction}
A number of long baseline interferometric gravitational wave detectors are working
around the world: in the USA (LIGO project), in Italy (VIRGO project, a joint French-Italian
collaboration), in Germany (GEO600 project), and in Japan (TAMA300 project). A network
of resonant bar detectors continues its operation: in the USA (ALLEGRO detector) and
in Italy (detectors AURIGA, NAUTILUS and EXPLORER (located in CERN near Geneva)).
These detectors are collecting a large amount of data that are being currently analyzed.
There is a proposed space borne detector LISA to be launched by NASA and ESA in the next
decade. The quest for gravitational waves in the data requires optimal
statistical methods and efficient numerical algorithms to search over very
large parameter spaces \cite{cutler-93,brady-98}. A standard method is the maximum likelihood detection method, which consists of searching for local maxima of the likelihood function with respect to the parameters \cite{davis-89}.
Assuming Gaussian noise in the detector, the maximum likelihood method consists of
correlating the data with templates defined over the parameter space.
In this paper we introduce a new tool for data analysis - a distance between the probability
density functions. This distance can be used to define the covering radius to design
an optimal (with smallest number of nodes) grid in the parameter space. One can use the
distance to determine the quality of the search templates - simplified models of the signal
over a reduced parameter space. The distance can also be used to study the problem of signal resolution - a problem that occurs in the estimation of parameters of white dwarf binary systems
in the data of planned space-borne detector LISA. This distance would play the same
role in gravitational wave data analysis as the line element defined by the Fisher matrix
interpreted as a Riemannian metric. However Fisher matrix is obtained as a Taylor expansion
up to the second order term of the Kullback-Leibler divergence and therefore is only an approximation.
Moreover the distance we introduce fulfills triangle inequality which is not true for Kullback-Leibler
divergence from which the Fisher matrix is obtained. Moreover our distance can be defined for any
probability densities, not only smooth and not only absolutely continuous with respect to each other.

In Section 2 we shall briefly review the problem of signal detection and parameter estimation.
In Section 3 we shall motivate and introduce the $L_1$-norm distance between two probability
density functions. We show that as a consequence of the triangle inequality, our distance is
an appropriate tool in several data analysis problems. In Section 4 we shall review the Kullback-Leibler divergence. In Section 5 we shall calculate the $L_1$-norm and the Kullback-Leibler divergence in several cases
important for applications, namely for the case of Gaussian probability density functions.
 In Section 6 we shall discuss applications of the $L_1$-norm to the problem of signal
 resolution, template placement and search template design for the simple case of
 a monochromatic signal. Section 7 concludes our paper.

\section{Problem of signal detection and parameter estimation}
\subsection{Signal detection and parameter estimation in Gaussian noise}
Suppose that we want to detect a known signal $s$ embedded in noise $n$.
The signal detection problem can be posed as a hypothesis testing problem,
where the null hypothesis is that the signal is absent, and the alternative hypothesis
is that the signal is present. A solution to this problem has been found by Neyman and Pearson
\cite{neyman-33}. They have shown that, subject to a given false alarm probability, the test
that maximizes the detection probability is the likelihood ratio test.
Assuming that the noise is {\em additive}, the data time series $x(t)$ can be written as
\be
\label{eq:D}
x(t) = n(t) + s(t).
\ee
In addition if the noise is
a {\em zero-mean}, {\em stationary}, and {\em Gaussian} random
process, the log likelihood function is given by
\be
\label{eq:CM}
\log\Lambda = (x|s) - \frac{1}{2}(s|s),
\ee
where the scalar product $(\,\cdot\,|\,\cdot\,)$ is defined by
\be
\label{eq:SP}
(x|y) := 4 \Re \int^\infty_0
\frac{\tilde{x}(f)\tilde{y}^{*}(f)}{\tilde{S}(f)}\,{\mathrm{d}}f.
\ee
In Eq.\ (\ref{eq:SP}) $\Re$ denotes the real part of a complex
expression, the tilde denotes the Fourier transform, the asterisk is
complex conjugation, and $\tilde{S}$ is the {\em one-sided
spectral density of the noise} in the detector.
Equation (\ref{eq:CM}) is called the {\em Cameron-Martin formula} \cite{davis-89}.
From the Cameron-Martin formula we immediately see that the in the Gaussian case, the likelihood ratio
test consists of correlating the data $x$ with the signal $s$ that is present in
the noise and comparing the correlation to a threshold.  Such a correlation
$G = (x|s)$ is called the {\em matched filter}.  The matched filter is a linear operation on
the data.

An important quantity is the optimal {\em signal-to-noise ratio}
$\rho$ defined by
\be
\label{eq:d} \rho^2 := (s|s) = 4 \Re \int^\infty_0
\frac{|\tilde{s}(f)|^2}{\tilde{S}(f)}\,{\mathrm{d}}f.
\ee

Since data $x$  are
Gaussian and $G$ is linear in $x$, it has a normal probability
density function. Probability  density distributions $p_0$ and $p_1$ of
correlation $G$ when respectively signal is absent and present are given by.
\begin{eqnarray}
  p_0(G) &=& \frac{1}{\sqrt{2\pi \rho^2}} \exp[-\frac{1}{2}\frac{G^2}{\rho^2}] \\
  p_1(G) &=& \frac{1}{\sqrt{2\pi \rho^2}} \exp[-\frac{1}{2}\frac{(G -
  \rho^2)^2}{\rho^2}].
\end{eqnarray}
Probability of false alarm $Q_F$ and of detection $Q_D$ are readily
expressed in terms of error functions.
\begin{eqnarray}
\label{eq:FA}
  Q_F &=& \frac{1}{2}[1 - erf(\frac{1}{\sqrt{2}}\frac{G_o}{\rho})] \\
\label{eq:PD}
  Q_D &=& \frac{1}{2}[1 - erf(\frac{1}{\sqrt{2}}(\frac{G_o}{\rho} - \rho))],
\end{eqnarray}
where $G_o$ is the threshold and the error function $erf$ is defined as
\begin{equation}
\label{eq:erf}
erf(x) = \frac{2}{\sqrt{\pi}}\int^x_0e^{-t^2}\,dt.
\end{equation}
Thus to detect the signal we proceed as follows. We choose a certain value of
the false alarm probability. From Eq.\,(\ref{eq:FA}) we calculate the threshold
$G_o$. We evaluate the correlation $G$. If $G$ is larger than the threshold
$G_o$ we say that the signal is present.
We see that in the Gaussian case, a single parameter --
signal-to-noise ratio $\rho$ determines both probabilities -- of false
alarm and detection, and consequently the receiver's operating
characteristic. For a given false alarm probability, the greater the
signal-to-noise ratio, the greater the probability of detection of
the signal.

In general, we know the signal as a function of several unknown parameters
$\theta$. Thus to detect the signal we also need to estimate its parameters.
A convenient method is the maximum likelihood method, by which
estimators are those values of the parameters that maximize
the likelihood ratio. Thus the {\em maximum likelihood estimators}
$\hat{\theta}$ of parameters  $\theta$ are obtained by solving the
set of equations
\begin{equation}
\frac{\partial\Lambda(\theta,x)}{\partial\theta_i} = 0,
\end{equation}
where $\theta_i$ is the $i$th parameter.
The quality of any parameter estimation method can be assessed using the
{\em Fisher information matrix} $\Gamma$ and the Cram\`er - Rao bound \cite{cramer-46}.
The components of this matrix are defined by
\begin{equation}
\label{eq:GA}
\Gamma_{ij}
:= {\mathrm E} \bigg[ \frac{\partial\log\Lambda}{\partial\theta_i}
\frac{\partial\log\Lambda}{\partial\theta_j} \bigg]
= -{\mathrm E} \bigg[ \frac{\partial^2\log\Lambda}{\partial\theta_i\partial\theta_j} \bigg].
\end{equation}
The Cram\`er-Rao bound states that for {\em unbiased} estimators,
the covariance matrix of the estimators $C\geq\Gamma^{-1}$.
(The inequality $A \geq B$ for matrices means that the matrix $A-B$
is nonnegative definite.) In the case of Gaussian noise, the formula for
the Fisher matrix takes the form
\begin{equation}
\label{eq:GAGA}
\Gamma_{ij} = (\frac{\partial s(\theta)}{\partial\theta_i}|\frac{\partial s(\theta)}{\partial\theta_j}),
\end{equation}
where the scalar product $(\,\cdot\,|\,\cdot\,)$ is given by Eq.\,(\ref{eq:SP}).
\subsection{The case of a monochromatic signal}
Let us consider an application of the maximum likelihood estimation method
to the case of a simple signal - a {\em monochromatic signal}.
The monochromatic signal depends on three parameters: amplitude $A_o$,
phase $\phi_o$, and angular frequency $\omega_o$, and it has the form
\begin{equation}
\label{eq:mon}
s = A_o\cos(\omega_o t - \phi_o).
\end{equation}
Let us rewrite the signal (\ref{eq:mon}) as
\begin{equation}
\label{eq:monl}
s = A_c\cos(\omega_o t) + A_s\sin(\omega_o t),
\end{equation}
where
\begin{eqnarray}
A_c &=& A_o\cos\phi_o, \\
A_s &=& A_o\sin\phi_o.
\end{eqnarray}
Using Parseval's theorem and assuming that the observation time $T$
is much longer than the period $2\pi/\omega_o$, we have
\begin{eqnarray}
(\cos(\omega_o t)|\cos(\omega_o t)) &\simeq& (\sin(\omega_o t)|\sin(\omega_o t))
\simeq \frac{T}{S_o}, \\
(\cos(\omega_o t)|\sin(\omega_o t)) &\simeq& 0,
\end{eqnarray}
where $S_o$ is the one-sided spectral density of noise of the detector
at frequency $\omega_o$.
Thus the log likelihood ratio is approximately given by
\begin{equation}
\log\Lambda = 2\frac{T}{S_o}[A_c <x\cos(\omega_o t)> + A_s <x\sin(\omega_o t)>
- \frac{1}{2}(A_c^2 + A_s^2)],
\end{equation}
where the operator $<\cdot>$ is defined as
\be
\label{eq:A}
<g(t)> = \frac{1}{T}\int^T_0g(t)\,dt = \int^1_0g(z T)\,dz,
\ee
where the last equation follows by introducing a dimensionless
time variable $z = t/T$.
The maximum likelihood estimators of $\hat{A}_c$ and $\hat{A}_s$
amplitudes $A_c$ and $A_s$ can
be obtained in a closed analytic form by solving the set of the following
two linear equations:
\begin{eqnarray}
<x\cos(\omega_o t)> - A_c = 0, \\
<x\cos(\omega_o t)> - A_s = 0.
\end{eqnarray}
We have
\begin{eqnarray}
\label{eq:mla}
\hat{A}_c = <x\cos(\omega_o t)>, \\ \nonumber
\hat{A}_s = <x\sin(\omega_o t)>.
\end{eqnarray}
By substituting the maximum likelihood estimators of amplitudes
into the log likelihood ratio we get
\begin{equation}
\log\Lambda_r = \frac{T}{S_o}[ <x\cos(\omega_o t)>^2 + <x\sin(\omega_o t)>^2].
\end{equation}
We shall denote the reduced likelihood ratio $\log\Lambda_r$
by ${\mathcal F}$ and we shall call it the ${\mathcal F}$ - statistic.
The maximum likelihood estimators $\hat{\phi}_o$ and
$\hat{A}_o$ of the phase and amplitude are given by
\begin{eqnarray}
\label{eq:mlp}
\hat{\phi}_o &=& \mbox{atan}[\frac{<x\sin(\omega_o t)>}{<x\cos(\omega_o t)>}], \\
\hat{A}_o &=& \sqrt{<x\cos(\omega_o t)>^2 + <x\sin(\omega_o t)^2>}.
\end{eqnarray}
Thus to find the maximum likelihood estimators of parameters of the
monochromatic signal, we first find the maximum of the ${\mathcal F}$ - statistic
with respect to angular frequency, and the angular frequency
$\hat{\omega}_o$ corresponding to the maximum of ${\mathcal F}$
is the maximum likelihood estimator of $\omega_o$.
Then we use Eqs.\,(\ref{eq:mlp}) with $\omega_o =\hat{\omega}_o$
to find the maximum likelihood estimators of phase and
amplitude. The maximum likelihood detection method consists of
correlating the data $x$ with two filters $F_c = \cos(\omega_o t)$
and $F_s = \sin(\omega_o t)$. We easily see that the ${\mathcal F}$ - statistic is invariant with respect to the following transformation
of the filters
\begin{eqnarray}
\sin(\omega_o t) \rightarrow  A_F \sin(\omega_o t + \phi_F), \\
\cos(\omega_o t) \rightarrow  A_F \cos(\omega_o t + \phi_F),
\end{eqnarray}
where $A_F$ and $\phi_F$ are arbitrary constants.

With the above approximations, the signal-to-noise ratio $\rho$ and the Fisher
matrix $\Gamma$ for the signal (\ref{eq:mon}) are given by
\begin{eqnarray}
\label{eq:snr}
\rho^2 &=& A_o^2\frac{T}{S_o}, \\
\label{FM}
\Gamma_{ij} &=&
\rho^2 \left(%
\begin{array}{ccc}
  \frac{1}{A_o^2} &     0           &  0            \\
  0               &     1           &  \frac{T}{2}  \\
  0               & \frac{T}{2}     &  \frac{T^2}{3}  \\
\end{array}%
\right),
\end{eqnarray}
where $i,j = (A_o, \phi_o, \omega_o)$.

\section{$L_1$-norm distance}
\subsection{Motivation}

Let $I$ denote a probabilistic space of events $i\in I$; for the sake of clarity
we will consider $I$ to be a set of finitely many elements in this initial
discussion (generalization to continuous probabilistic spaces and measures will
be commented upon later in this section). Let $p(i)$, $q(i)$ be two
different probability distributions (strictly speaking: probabilistic measures)
defined on $I$.

We propose to define the distance $d_L$ between two probability distributions
$p$ and $q$ as
\be
d_L = \frac{1}{2}\sum_{i\in I}\left|p(i) - q(i)\right|
\ee
where the sum is taken over the whole event space $I$.

The space of probability distributions with the above distance is a
metric space because the distance fulfills all the axioms of a metric
i.e. $d(p,q) = 0$ if and only if $p = q$, $d(p,q) = d(q,p$)  and the triangle
equality holds: $d(p,q) + d(q,r) \geq d(p,r)$.

Remark: $d_L(p,q)\leq 1$ for any $p$, $q$; the maximal value is achieved
if and only if $p$ and $q$ have disjoint supports, i.e. when for every $i\in I$ either
$p=0$ or $q=0$.

We motivate the above definition by the following example.

Consider the following situation: we are performing an observation
or measurement on a system where either of two random processes may
be operating at a given time. To our best knowledge, each of the two
processes is described by a determined probability distribution of
the measured results; moreover, we conjecture (or know) the probabilities
that each of these processes is operating in a given instance of the
measurement or observation -- alternatively, if such knowledge is
lacking, we assume that each of the two is equally likely to be in
effect. In every instance of the experiment, we need to decide, based
on the obtained result, which of the two processes was more likely
to be operating, and we need to somehow estimate our overall confidence
in these decisions, based on how much the two pre-determined probability
distributions differ over the space of observed results.

To be more precise, let $I$ be the space of possible outcomes
of our measurements, and let $p(i),$ $q(i)$ $(i\in I)$ be the probabilities
that the result $i$ is obtained when the process ${\cal \mathcal{P}}$
(respectively, $\mathcal{Q}$) is in operation. Recall that we are regarding
$I$ to be a finite set, as is in fact the case in many realistic
experimental setups (even if the number of elements of $I$ is typically
quite large).

We consider the special case when both processes are equally likely,
a natural assumption when we lack any \emph{a priori} knowledge about the probability
of either process being in operation for a given event.

Clearly, the total probability that a measurement will produce the
outcome $i\in I$ is given by the combination

\[
\frac{1}{2}(p(i) + q(i)),\]

while the conditional probability that, assuming the outcome $i_{0}\in I$
was obtained, it resulted from the process $\mathcal{P}$, is

\[
\frac{p(i_{0})}{p(i_{0}) + q(i_{0})}\]
 and likewise for process $\mathcal{Q}$:

\[
\frac{q(i_{0})}{p(i_{0}) + q(i_{0})}.\]

We see that the relative likelihood of each process being in operation,
given a specific outcome, leads to basing our decision on the partition
of the space $I$ into two subsets

\[
\mathbf{P}=\{ i\in I:\: p(i)>q(i)\}\]

and

\[
\mathbf{Q}=\{ i\in I:\: q(i)>p(i)\},\]

where if the outcome is $i_{0}\in\mathbf{P}$ it is more likely to
result from process $\mathcal{P}$, and correspondingly for $i_{0}\in\mathbf{Q}$
and $\mathcal{Q}$.

Note that we have neglected here the subset of $I$ where $p(i)=q(i)$;
i.e., where both processes were both likely to be operating. For such
outcomes the decision is arbitrary and we will see below that it has
no impact upon our conclusions.

Now, in every case (for any value of $i_{0}\in I$) the greater likelihood
rule stated above might lead us to err; the probability that a decision
in favor of the more likely of either $\mathcal{P}$ or $\mathcal{Q}$
is mistaken, conditional upon the outcome being $i_{0}$, is given
by
\[
PE(i_{0})=\frac{q(i_{0})}{p(i_{0})+q(i_{0})}\;(i_{0}\in\mathbf{P})\]
 and

\[
PE(i_{0})=\frac{p(i_{0})}{p(i_{0})+q(i_{0})}\;(i_{0}\in\mathbf{Q}).\]
Under our assumptions, the overall probability that the greater likelihood
rule will fail is therefore obtained by summing the above, weighted by the
probability of the result being $i_0$, over all of $I$:

\[
PE=\frac{1}{2}({\displaystyle \sum_{i\in\mathbf{P}}q(i)+\sum_{i\in\mathbf{Q}}p(i)}).\]

For simplicity we ignore here the fact that this might need to be corrected
by adding $1/2$ times the probability of hitting the subset of $I$
where $p(i)=q(i)$ (whatever the arbitrary rule applied to decide
for such outcomes, there is a probability of $1/2$ that it is wrong).

Note that the overall probability of error $PE$ given above may be,
in the extreme case, equal to zero -- this, when the probability distributions
$p(i)$ and $q(i)$ have disjoint supports, or (in the other extreme)
to $1/2$ -- this is obtained when both distributions are identical.

The above formula may be simplified to
\[
PE=\frac{1}{2}\sum_{i\in I}\min(p(i),q(i)).\]

Note that this form already takes correctly into account the possibility
that $p(i)=q(i)$ for some values of $i$.

Further simplification follows by using
\[
\min(p(i),q(i))=\frac{1}{2}(p(i)+q(i)-\left|p(i)-q(i)\right|)\]

which leads to

\[
PE=\frac{1}{4}\sum_{i\in I}(p(i)+q(i)-\left|p(i)-q(i)\right|)\]

and, ultimately, to

\[
PE=\frac{1}{2}-\frac{1}{4}\sum_{i\in I}\left|p(i)-q(i)\right|\]

i.e.

\be
\label{eq:PE}
PE=\frac{1}{2}(1 - d_L(p,q)).
\ee

Obviously, when $p=q$, and consequently, $d_L(p,q)=0$, $PE=1/2$ --
expressing the fact that it is impossible to distinguish between two
processes whose experimental outcomes are identical. The other extreme
is achieved when $PE=0$: this follows when $d_L(p,q)=1$, which, as
remarked before, happens if and only if $p$ and $q$ have disjoint supports. In other
words, when every possible result of measurement can be produced by
only one of the two processes under consideration, there can be no mistake
in determining which of those processes was operating.

From the above we see that our definition of the $L_1$-norm distance
between probability distributions admits a natural probabilistic
interpretation: the $L_1$-norm distance is closely related to the
level of reliability of the rule of greater likelihood, applied to
determining which of two \emph{a priori} equally likely processes is
being observed in an experiment.

\subsection{Continuous probabilistic space}
The above discussion, beginning with the definition of the
$L_1$-norm distance, may be generalized to the case when $I$ is a
continuous probabilistic space, with the obvious substitutions of
sums by integrals etc. To perform this generalization rigorously, it
is simplest to assume the existence of a ''reference'' measure $\mu$
on $I$ such that both probabilistic measures involved in the
definition of $d_L$ are absolutely continuous with respect to $\mu$,
and may therefore be represented by non-negative functions $p(i)$,
$q(i)$. The distance $d_L(p,q)$ is then given by
\[
d_L(p,q)=\frac{1}{2}\int_I\left|p(i)-q(i)\right|d\mu(i).
\]
It then remains to be shown that the result is independent of the choice of
$\mu$. This easily follows by observing that any other $\mu_1$ fulfilling the
required properties must be absolutely continuous with respect to $\mu$, and
vice versa, when restricted to the sum of the supports of $p$ and $q$:
\[
\mu_1 = \rho\mu \ \{ i\in \mathrm{supp}(p)\cup \mathrm{supp}(q) \}
\]
with $\rho$ a strictly positive function, and
\[
p_1(i)=p(i)/\rho(i),
\]
and likewise for $q$. Clearly, $d_L$ is independent of $\rho$.

In fact, it follows from classical work by Riesz on measure theory that it is not
even necessary to assume the existence of a ''reference'' measure $\mu$, as our
definition of $d_L$ is simply a special case of Riesz's definition of the
norm on the space of additive functions of sets.

Using the fact that probability density function is non-negative we can write
the distance $d_L$ as
\begin{equation}
d_L = \frac{1}{2}\int_X\ q \left|\frac{p}{q} - 1 \right|\,dx = \frac{1}{2}E_q[\left|\Lambda - 1\right|]\,dx,
\end{equation}
where $\Lambda = \frac{p}{q}$ is the likelihood ratio.

\subsection{Non-uniform priors}
It is not difficult to generalize the discussion to the case when the two processes,
described by probability distributions $p(i)$ and $q(i)$, are no longer treated
symmetrically. When the assumption od equal \emph{a priori} probabilities
is relaxed, the total probability of outcome $i$ is now given by
\[
Pp(i) + Q q(i)
\]
with $P + Q = 1$. Let us now think in terms of searching for some ''signal'', whose
presence leads to measurements distributed according to $p(i)$, versus the
''pure noise'' described by $q(i)$.  The signal is more likely than not to be present for
outcomes that fall within the set
\[
\mathcal{P} = \{i\in I:\: Pp(i) > Q q(i)\}.
\]
As long as we can assume that $q(i)$ is nonzero for all $i\in I$, this inequality
may be written in terms of the likelihood ratio $p/q$:
\[
\frac{p(i)}{q(i)} > k
\]
where the ''detection threshold'' $k$ is given in terms of the \emph{a priori}
probabilities $P$ and $Q$:
\[
k = \frac{Q}{P} = \frac{1-P}{P},
\]
and conversely
\[
P = \frac{1}{k+1} ,\, Q = \frac{k}{k+1}.
\]
It is usual in signal detection theory to consider separately the
,,false alarm probability'' (the probability that a decision rule
mistakenly leads us to believe the signal to be present), and the
,,false dismissal probability'' (that the signal might be missed
when it is in fact present). Instead, we restrict ourselves here to
considering, as before, the total probability of an erroneous
decision $PE$. In the current model, $PE$ is given by:
\[
PE = \sum_{i\in I} \mathrm{min}(Pp(i), Qq(i)) =
\frac{1}{2}\left( 1 - \sum_{i\in I} |Pp(i) - Qq(i)|\right).
\]
While this can no longer be expressed in terms of $d_L(p, q)$, contrary
to the case of equal \emph{a priori} probabilities, the expression above
still involves the $L_1$-norm of the difference between the two (non-normalized)
densities (measures) $Pp$ and $Qq$.

Let us now consider two different signals, described by distributions $p_1$ and
$p_2$, which are close to each other in the sense of the $L_1$ distance (\emph{i.e.}
$d_L(p_1, p_2) < \eps$ for some small $\eps$), and a pure noise described by $q$.
A simple application of the triangle
inequality for the $L_1$-norm leads to the inequality
\[
\sum_{I}|Pp_2 - Qq| \leq \frac{2\eps}{k + 1} + \sum_{I}|Pp_1 - Qq|.
\]
Note that in the case of equal a priori probabilities the above simply
reduces to the triangle inequality:
\begin{equation}
\label{eq:BINE}
d_L(p_2,q) \leq \eps + d_L(p_1,q).
\end{equation}
In other words: when two signals (or rather, their corresponding data distributions)
differ by no more than $\eps$ in term of the $L_1$ distance, the probabilities
of confusing presence of the signal with pure noise (under equal detection threshold)
differ by a term which is as well of order $\eps$. This property makes the distance
$d_L$ appropriate for definition of covering radius in the construction of the grid of templates. When probability distribution $p_1$ corresponds to the true signal and $p_2$
to the signal with parameters at the node of the grid the inequality (\ref{eq:BINE}) tells
us that the error probability will not increase by more than $\epsilon$.
Likewise in the problem of designing suboptimal templates that approximate the true
signal the inequality (\ref{eq:BINE}) says that when distance $d_L$ of the signal to the template is less than $\epsilon$ the probability of error does not increase by more than
$\epsilon$.

\section{Kullback-Leibler divergence}

A useful measure of distance between the two probability measures
$p(x)$ and $q(x)$ was defined by Kullback and Leibler \cite{kull-51}.
The Kullback - Leibler divergence $d_{KL}(p,q)$ is defined as
\begin{equation}
\label{eq:KLd}
    d_{KL}(p,q) = E_p[\log\frac{p}{q}] + E_q[\log\frac{q}{p}].
\end{equation}
Using the likelihood ratio $\Lambda = \frac{p}{q}$ we can write the
Kullback - Leibler divergence as
\begin{equation}
\label{eq:KLdL}
    d_{KL} = E_p[\log\Lambda] - E_q[\log\Lambda].
\end{equation}
In Bayesian statistics the KL divergence can be used as a measure of the "distance" between the prior distribution and the posterior distribution. In coding theory, the KL divergence can be interpreted as the needed extra message-length per datum for sending messages distributed as q, if the messages are encoded using a code that is optimal for distribution p.

\section{Examples}

In this Section we shall calculate the distance $d_L$ and the
Kullback-Leibler divergence $d_{KL}$ in several cases useful for
applications.

\subsection{$d_L$ distance}
\subsubsection{Gaussian probability density function.}
Let $p$ and $q$ be Gaussian probability density functions
with mean $\mu$ and $\nu$ respectively
and the same variance $\sigma^2$ then the distance $d_L$ is given by
\begin{equation}
\label{eq:WRgo}
d_L = erf[\frac{1}{2\sqrt{2}}\frac{|\mu - \nu|}{\sigma}],
\end{equation}
where $erf$ is the error function defined by Eq.\,(\ref{eq:erf}).

For the case of two arbitrary Gaussian probability density functions $p$
and $q$ with means $\mu_p$, $\mu_q$ and variances $\sigma^2_p$, $\sigma^2_q$
respectively we have
\begin{eqnarray}
d_L = \frac{1}{2}\left\{erf[\frac{1}{\sqrt{2}}\frac{x_1 - \mu_p}{\sigma_p}] -  erf[\frac{1}{\sqrt{2}}\frac{x_2 - \mu_p}{\sigma_p}]
      - (erf[\frac{1}{\sqrt{2}}\frac{x_1 - \mu_q}{\sigma_q}] -  erf[\frac{1}{\sqrt{2}}\frac{x_2 - \mu_q}{\sigma_q}])\right\},
\end{eqnarray}
where
\begin{eqnarray}
x_{1,2} &=& \frac{-b \pm \sqrt{b^2 - 4 ac}}{2 a}, \\
a &=& \frac{1}{\sigma_p^2} - \frac{1}{\sigma_q^2}, \\
b &=& -2 (\frac{\mu_p}{\sigma_p^2} - \frac{\mu_q}{\sigma_q^2}),\\
c &=& \frac{\mu_p^2}{\sigma_p^2} - \frac{\mu_q^2}{\sigma_q^2} - 2\ln\frac{\sigma_q}{\sigma_p}.
\end{eqnarray}

Let $p({\bf x})$ and $q({\bf x})$ be two $n-$dimensional multivariate Gaussian
probability density functions with vector means ${\bf \mu}$ and ${\bf \nu}$ respectively
and the same covariance matrix ${\bf \Sigma}$.
Thus $p({\bf x})$ is given by
\begin{equation}
p({\bf x}) = \frac{1}{(2\pi)^{n/2} \sqrt{det {\bf \Sigma}}}
\exp[-\frac{1}{2}{({\bf x} - \bf \mu})'{\bf \Sigma}^{-1}({\bf x} - {\bf \mu})],
\end{equation}
where $'$ denotes transpose and similarly $q({\bf x})$.
To calculate the distance $d_L$ we first
perform a change of variables so that the covariance matrix ${\bf
\Sigma}$ is diagonal and the diagonal elements are equal. We can
then rotate the vector ${\bf }$ so that it is aligned along the
$x_1$ axis. These transformations bring the calculation of distance
$d_L(p,q)$ to one dimensional case. Consequently we have
\begin{equation}
\label{eq:WRgon}
d_L = erf[\frac{1}{2\sqrt{2}}\sqrt{({\bf \mu} - {\bf \nu})'{\bf \Sigma}^{-1}({\bf \mu} - {\bf \nu})}].
\end{equation}

\subsubsection{Stationary Gaussian process.}
Let us consider the case of signals $s_1$ and $s_2$ added to a stationary Gaussian random process.
We then have two Gaussian probability density $p_1$ or $p_2$ when respectively
the signal $s_1$ or the signal $s_2$ is present. We can obtain the distance $d_L(p_1,p_2)$
as the limit of the case of the multivariate Gaussian distributions by
replacing $({\bf \mu} - {\bf \nu})'{\bf \Sigma}^{-1}({\bf \mu} - {\bf \nu})$
with $(s_1 - s_2|s_1 - s_2)$ where the scalar product $(\,\cdot\,|\,\cdot\,)$ is defined by
Eq.\,\ref{eq:SP}. Thus in this case we have
\begin{equation}
\label{eq:WRgoc}
d_L(s_1,s_2) = erf[\frac{1}{2\sqrt{2}}\sqrt{(s_1 - s_2|s_1 - s_2)}].
\end{equation}
We have introduced a notation
\be
d_L(s_1,s_2) = d_L(p_1,p_2),
\ee
where $p_1, p_2$ are probability density distributions when signal
$s_1, s_2$ respectively are present in the data.
In the case of detection of signal in noise when we have two Gaussian
probability density functions $p_1$ when the signal $s$ is present and $p_0$
when the signal is absent the distance $d_L(p_0,p_1)$ is immediately
obtained form Eq.\,(\ref{eq:WRgoc}) above:
\begin{equation}
d_L = erf[\frac{1}{2\sqrt{2}}\sqrt{(s|s)}].
\end{equation}

Finally suppose that $p$ and $q$ belong to the same family
$p_{\theta}(x)$ of probability density functions parameterized by
parameter $\theta$. Let $p(x) = p_{\theta}(x)$ and $q(x) = p_{\theta
+ \delta\theta}(x)$ where $\delta\theta$ small. Then using Taylor expansion
to the first order we have
\be
d_L(\theta,\theta + \delta\theta) =
 \int_X\left|\frac{\partial p}{\partial\theta_i}\delta\theta_i\right|\,dx,
\ee
where we have introduced a short hand notation
\be
d_L(\theta,\theta') = d_L(p_{\theta},p_{\theta'}).
\ee

\subsection{$d_{KL}$ divergence}
\subsubsection{Gaussian probability density function.}
For the case of two Gaussian probability density functions
with means $\mu_p$, $\mu_q$ and variances $\sigma^2_p$, $\sigma^2_q$
respectively we have
\begin{equation}
d_{KL} =
\frac{1}{2}\frac{(\sigma_p^2  - \sigma_q^2)^2 +
(\sigma_p^2 + \sigma_q^2)(\mu_p - \mu_q)^2}{\sigma_p^2 \sigma_q^2}
\end{equation}
When the two Gaussian probability distributions have the same variance equal to $\sigma^2$
the above formula reduces to
\begin{equation}
\label{eq:KLgo}
d_{KL} = \left(\frac{\mu_p - \mu_q}{\sigma}\right)^2.
\end{equation}
For the case of $n-$dimensional multivariate Gaussian probability density
functions $p$ and $q$ the divergence $d_{KL}(p,q)$ is given by
\begin{equation}
d_{KL}(p,q) = ({\bf \mu} - {\bf \nu})'{\bf \Sigma}^{-1}({\bf \mu} - {\bf \nu}).
\end{equation}

\subsubsection{Stationary Gaussian process.}
In the case of detection of signal in Gaussian stationary noise we immediately
obtain the Kullback-Leibler divergence $d_{KL}$ between the probability
density functions $p_0$ and $p_1$ when respectively signal is absent and present
using the Cameron-Martin formula (\ref{eq:CM}). In this case we have
\be
\label{eq:KLdS}
d_{KL} = (s|s).
\ee
Thus in this case the Kullback-Leibler divergence is precisely equal to the
signal-to-noise ratio square.
Suppose that $p$ and $q$ belong to the same family
$p_{\theta}(x)$ of probability density functions parameterized by
parameter $\theta$. Let $p(x) = p_{\theta}(x)$ and $q(x) = p_{\theta
+ \delta\theta}(x)$ where $\delta\theta$ small. Then one can show by
Taylor expansion that to the first order the KL-divergence
$d_{KL}(\theta,\theta + \delta\theta)$ between $p_{\theta}$ and
$p_{\theta + \delta\theta}$ is given by
\begin{equation}
\label{eq:KLF}
ds^2 = d_{KL}(\theta,\theta + \delta\theta) = \Gamma_{ij} \delta\theta_i
\delta\theta_j.
\end{equation}
From Eqs.\, (\ref{eq:KLdS}) and (\ref{eq:KLF}) we see that the Kullback-Leibler divergence
is directly related to the basic quantities used in detecting signals in noise
and estimating their parameters - the signal-to-noise ratio and the Fisher information matrix.
Also the equation (\ref{eq:KLF}) reinforces the interpretation of
the Fisher information matrix as a Riemannian metric on the
parameter space \cite{owen-96} as the square root of the Kullback-Leibler divergence of
probability density functions of closely spaced parameters is the
line element $ds$ for the Fisher metric $\Gamma$.

\section{Comparing the two distances}
In contrast to the distance $d_L$ the
Kullback-Leibler divergence is not a metric because it does not fulfill
the triangle inequality.
The distance $d_L$ has the advantage over the Kullback-Leibler divergence
that it exists even if the two probability measures are not absolutely continuous
with respect to each other. If the probability measure $p$ is not absolutely
continuous with respect to $q$ the divergence $d_{KL}$ does not exist.

\begin{figure}
  \includegraphics[width=13cm]{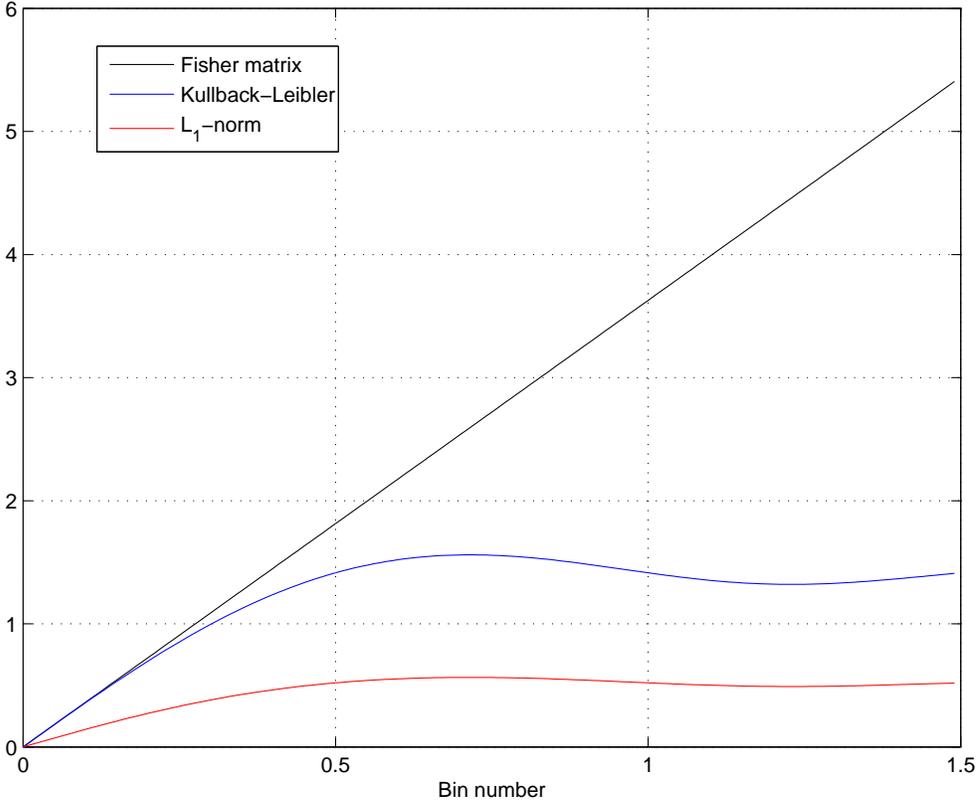}\\
  \caption{\label{fig:comp2} Comparison of the $L_1$-norm distance, the Kullback-Liebler divergence, and
  the line element defined by the Fisher matrix for the case of two monochromatic signals
  with the same amplitudes and phases but different frequencies as functions
  of frequency difference expressed in Fourier bins.}
\end{figure}

Let us compare the $L_1$-norm distance with the Kullback-Leibler divergence
and also with the line element $ds$ defined by the Fisher matrix
(Eq.\,\ref{eq:KLF}) for the case of a monochromatic signal.
Let us calculate the norm
$N = (s_1 - s_2|s_1 - s_2) = (s_1|s_1) + (s_2|s_2) - 2 (s_1|s_2)$
where $s_1$ and $s_2$ be two monochromatic signals with
amplitudes $A_1$ and $A_2$, phases $\phi_1$ and $\phi_2$ and angular frequencies
$\omega_1$ and $\omega_2$ respectively.
Assuming that over the bandwidth $[\omega_1 \,\,\, \omega_2]$ spectral density
is constant and equal to $S_o$, using Parseval's theorem,
and assuming that the observation time $T$
is much longer than the period $2\pi/\omega_o$ we have
\begin{eqnarray}
\label{eq:norm}
N =  \rho_1^2 + \rho_2^2 - 2\rho_1\rho_2
[<\cos(\Delta\omega t)>\cos(\Delta\phi) - <\sin(\Delta\omega t)>\sin(\Delta\phi)],
\end{eqnarray}
where $\rho_1$, $\rho_2$ are signal-to-noise ratios for
signals $s_1$ and $s_2$ respectively,
$\Delta\omega = \omega_1 - \omega_2$, $\Delta\phi = \phi_1 - \phi_2$.
The operator $<\cdot>$ is defined by Eq.\,(\ref{eq:A}).
The Kullback-Leibler divergence $d_{KL}(s_1,s_2)$ between the Gaussian probability density functions
$p_1$ and $p_2$ for signals $s_1$ and $s_2$ and the distance $d_L(s_1,s_2)$ are given by (see Eq.\,(\ref{eq:WRgoc})):
\begin{eqnarray}
d_{KL} &=& N,\\
\label{eq:dLgaus}
d_L &=& erf (\frac{1}{2\sqrt{2}}\sqrt{N}).
\end{eqnarray}
Let us assume that the two signals $s_1$ and $s_2$ have the same amplitudes and
phases. Then we have $\rho_1 = \rho_2 = \rho$ and the distance $d_L$ and the divergence $d_{KL}$  are given by
\begin{eqnarray}
d_{KL} &=& 2\rho^2(1 - \frac{\sin(\Delta\omega T)}{\Delta\omega T}),\\
d_L &=& erf (\frac{1}{2\sqrt{2}}\sqrt{d_{KL}}).
\end{eqnarray}
The line element $ds^2$ defined by the Fisher matrix is given by the first non-vanishing
term of the Taylor expansion of $d_{KL}$ in $\Delta\omega$.
\be
ds = \sqrt{\Gamma_{\omega_o\omega_o}(\Delta\omega)^2} = \frac{\rho \Delta\omega T}{\sqrt{3}}, \\
\ee
where $\Gamma_{\omega_o\omega_o}$ is the component of the Fisher matrix given by Eq.\,(\ref{FM}).
It is clear form Eqs.\,(\ref{eq:WRgo}) and (\ref{eq:KLgo}) that it is appropriate
to compare $L_1$-norm distance with a square root Kullback-Leibler divergence.
In Figure \ref{fig:comp2} we have plotted $ds, \sqrt{d_{KL}}$, and $d_L$
as functions of frequency difference expressed in Fourier bins.
A Fourier bin is equal to $1/T$. We see from Figure \ref{fig:comp2} that for frequency difference larger than a quarter of a Fourier bin the distance $ds$ based on the Fisher matrix begins to deviate substantially from the Kullback-Leibler divergence. This shows limitations of the applicability
of the Fisher matrix. This is a consequence of the fact that the Fisher matrix is obtained
by a Taylor expansion up to the second order terms of the Kullback-Leibler divergence.
As we shall see in the Section \ref{ssec:grid} below we can apply the distance measures to construction of a grid of templates by considering $s_1$ as a signal and
$s_2$ as a template.
Taylor expansion may be a good approximation for a very fine grid.
However for computationally intensive searches like a search for periodic sources
\cite{astone-02,papa-06} where one needs to choose a loose grid the Taylor expansion
of KL-divergence up to second order terms is not accurate.

\section{Applications}
Using the example of the monochromatic signal let us consider several applications of the
distance $d_L$ to problem of detection of signal in noise and estimation of parameters.

\subsection{Signal resolution}
\begin{figure}
  \includegraphics[width=13cm]{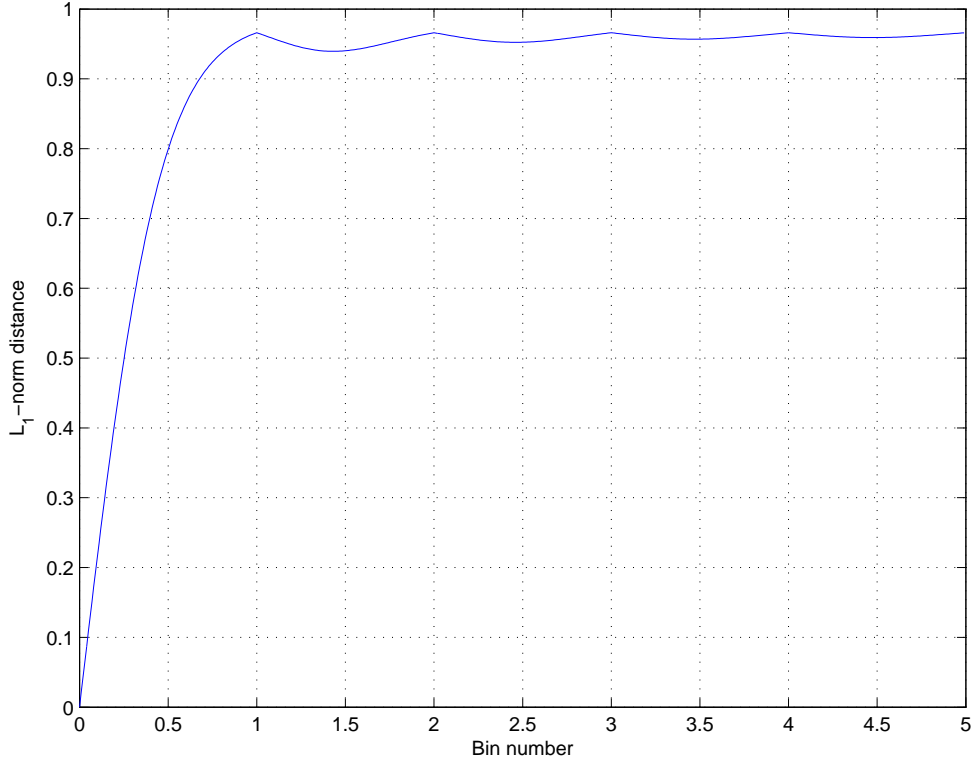}\\
  \caption{\label{fig:mon} $L_1$-norm distance between Gaussian probability density
  functions for two monochromatic signals
as a function of their frequency difference. We assume that the
signal-to-noise ratios of both signals are equal to 3.}
\end{figure}

\begin{figure}
  \includegraphics[width=13cm]{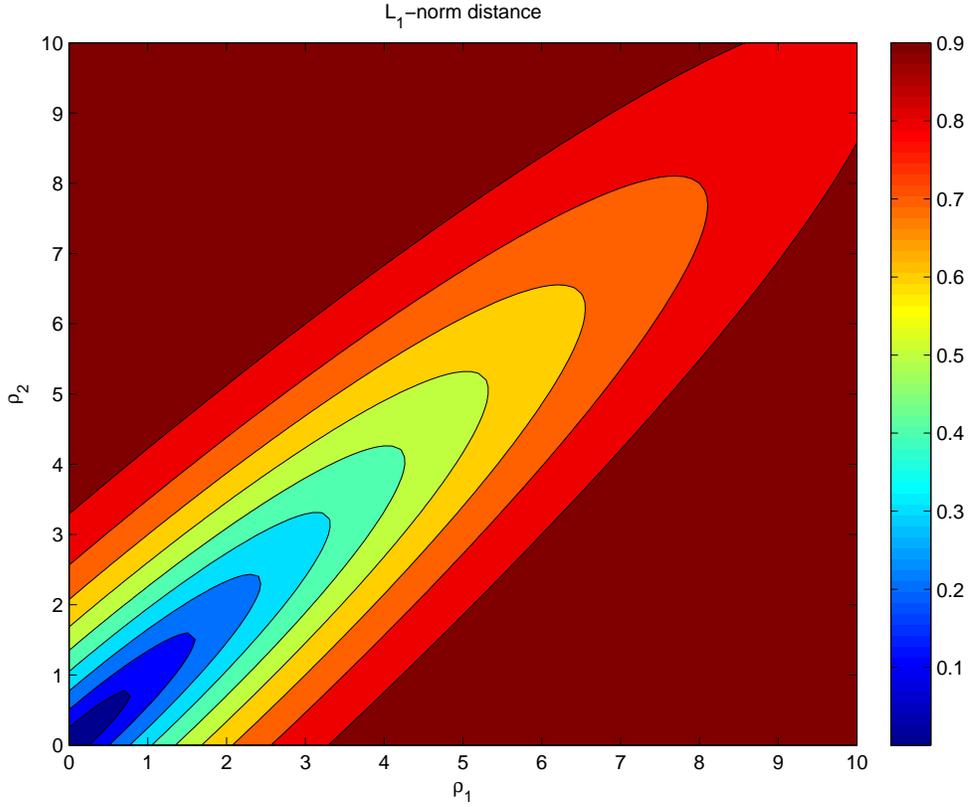}\\
  \caption{\label{fig:pe2} $L_1$-norm distance as a function of signal-to-noise ratios
$\rho_1$ and $\rho_2$ of two signals for a fixed difference between
the frequencies of the signal equal to one bin. }
\end{figure}

The distance $d_L(s_1,s_2)$ determines how well we can resolve two signals
$s_1$ and $s_2$. The larger the distance the better the signal resolution.
It is useful to have a reduced form of the distance that depends only on the frequencies
$\omega_1$ and $\omega_2$. We can achieve a reduction of the phase parameter by considering
the worst case i.e. the minimum of the distance $d_L$ given by Eqs.\, (\ref{eq:norm}) and
(\ref{eq:dLgaus}) over the phase difference $\Delta\phi$.
One easily finds an analytic formula for this minimum.
\begin{equation}
\label{eq:dLmin}
d_{Lmin} = erf\left[\frac{1}{2\sqrt{2}}\sqrt{\rho_1^2 + \rho_2^2 - 2\rho_1\rho_2
\sqrt{\frac{2(1 - \cos(\Delta\omega T))}{(\Delta\omega T)^2}}}\right].
\end{equation}
There is no nontrivial minimum with respect to amplitudes (the
minimum of $d_{Lmin}$ with respect to amplitudes $A_1$ and $A_2$ is zero).
In Figure \ref{fig:mon} we plot the
distance $d_L$ given by Eq.\,(\ref{eq:dLmin}) as a function of the
difference in angular frequencies $\Delta\omega$ expressed in
Fourier bins. We assume that
signal-to-noise ratios of the signals $s_1$ and $s_2$ are equal and
equal to 3.
We see that the distance increases as the difference
between the frequencies increases. We also see that there is a
substantial increase in the distance when the difference in
frequencies between the two signal becomes one bin. This characteristic
increase over one bin is independent of the signal-to-noise ratios
of the signals. This justifies a folk theorem that two monochromatic
signals can be resolved when their frequencies differ by one bin.
In Figure \ref{fig:pe2} we have plotted the distance as a function
of signal-to-noise ratios of the two signals for a fixed difference
between the frequencies of the signals equal to one bin.
We see that the distance $d_L$ increases as the signal-to-noise
ratio increases and also as the difference between the signal-to-noise ratios of the two
signals increases. We see that we can achieve an arbitrary large distance
and consequently arbitrary large resolvability of the signal if its signal-to-noise ratio is sufficiently large.

\subsection{Grid of templates}
\label{ssec:grid}
\begin{figure}
  \includegraphics[width=13cm]{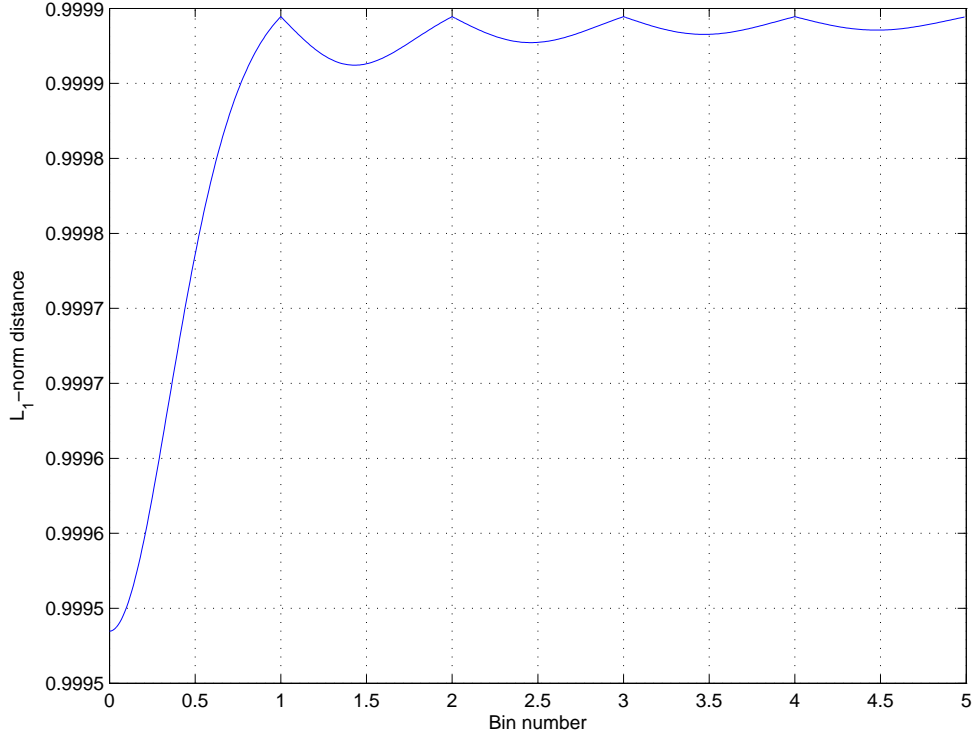}\\
  \caption{\label{fig:mon_grid}$L_1$-norm distance maximized over phase as a function of frequency spacing for
  a monochromatic signal. The signal-to-noise ratio is set to 8.}
\end{figure}

We can use the distance $d_L$ in construction of the grid of templates.
We construct the grid of templates in such a way that the maximum distance between
a signal and a filter is less than a certain specified value.
Let us first calculate the distance
$d_L(s,s_F)$ where $s$ is the signal with parameters $A_o, \phi_o, \omega_o$ and $s_F$ is the template. The template $s_F$ is just a monochromatic signal with some test parameters
$A_F, \phi_F, \omega_F$.
From Section 2 we know that the optimal statistic ${\mathcal F}$ is invariant with respect to transformation of the amplitude
and the phase of the filter and we can set the amplitude of the filter $A_F$ so that its signal-to-noise ratio
is equal to 1 and we can choose the phase of the filter $\phi_F$ to be equal to $0$.
With these simplifications the distance $d_L$ is given by
\begin{equation}
d_L =  erf\left\{\frac{1}{2\sqrt{2}}\sqrt{\rho^2 + 1 - 2\rho
[<\cos(\Delta\omega t)>\cos(\phi_o) - <\sin(\Delta\omega t)>\sin(\phi_o)}]\right\}.
\end{equation}
Since the ${\mathcal F}$-statistic depends only on the frequency
we only need the grid in the frequency space and consequently we
need to obtain a reduced distance that depends only on the
frequencies of the signal and filter. Again like in the case of
signal resolution it is natural to consider the worst case scenario.
In this case it corresponds to maximum of the distance $d_L$ with
respect to phase of the signal $\phi_o$. We easily get
\begin{equation}
\label{eq:dLgrid}
d_{Lmax} = erf\left[\frac{1}{2\sqrt{2}}\sqrt{\rho^2 + 1 + 2\rho
\sqrt{\frac{2(1 - \cos(\Delta\omega T))}{(\Delta\omega T)^2}}}\right].
\end{equation}
The function $d_{Lmax}$ for signal-to-noise ratio $\rho > 0$ is a monotonically
increasing function of $\rho$. We choose the grid
using the distance $d_L$ to determine the covering radius of the grid.
To calculate the covering radius we can set the signal-to-noise ratio
$\rho$ in Eq.\,(\ref{eq:dLgrid}) equal to the threshold value of $\rho$
used in the search. In Figure \ref{fig:mon_grid} we plot the distance
$d_{Lmax}$ as a function of the frequency difference  $\Delta\omega$ between
the signal and the filter for signal-to-noise equal to 8.


\subsection{Search templates}

\begin{figure}
  \includegraphics[width=13cm]{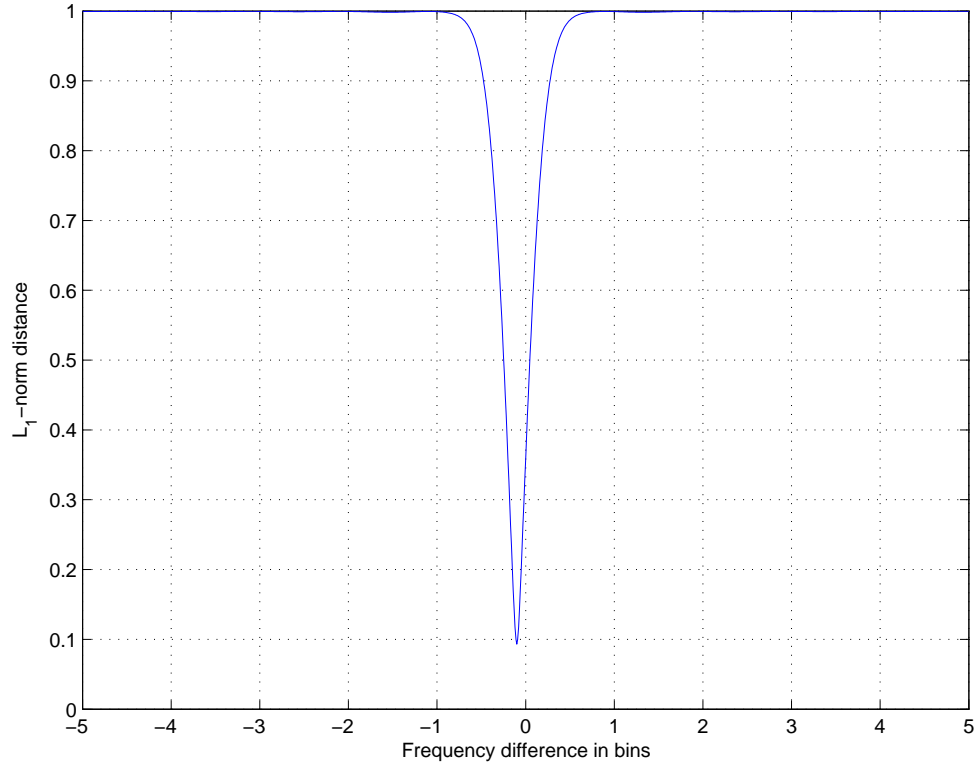}\\
  \caption{\label{fig:mon_tem} The $L_1$-norm distance between the frequency modulated signal
  and a monochromatic signal filter for frequency modulation parameter $\omega_1 = 2\pi/10/T^2$.}
\end{figure}

Very often we do not have the exact model of the signal that we are searching for.
Let us suppose that the signal that we expect to detect in noise is linearly
modulated in frequency and has the following form
\begin{equation}
\label{eq:mod}
s = A_o\cos(\omega_o t + \omega_1 t^2 + \phi_o).
\end{equation}
Let us also suppose that we know that $\omega_1$ is small and we can
expect to detect the signal (\ref{eq:mod}) with a monochromatic
signal template that has no frequency modulation.
The distance $d_L$ is given by
\begin{eqnarray}
d_L =  erf\left\{\frac{1}{2\sqrt{2}}\sqrt{\rho^2 + 1 - 2\rho
[S \cos(\Delta\phi) - C \sin(\Delta\phi)}]\right\},
\end{eqnarray}
where
\begin{eqnarray}
S = <\sin(\Delta\omega t + \omega_1 t^2)>, \\
C = <\cos(\Delta\omega t + \omega_1 t^2)>. \\
\end{eqnarray}
To determine the quality of the search template we need to find the
minimum of the distance with respect to parameters of the filter.
The minimum of $d_L$ with respect to phase $\phi_F$ of the filter
can easily be obtained and is given by
\begin{equation}
\label{eq:dLminTem}
d_{Lmin} = erf[\frac{1}{2\sqrt{2}}\sqrt{\rho^2 + \rho^2_F - 2\rho\rho_F\sqrt{S^2 + C^2}}].
\end{equation}
The minimum value $d_{Lmin}$ is independent of the phase of $\phi_o$ of the signal.
The minimum is a monotonic function of $\rho$.
Consequently to find the minimum of $d_L$ with respect to
the frequency parameter $\omega_F$ of the filter
is equivalent to find the maximum of the function
\begin{equation}
FF = \sqrt{S^2 + C^2}
\end{equation}
with respect to $\omega_F$.
In Figure \ref{fig:mon_tem} we have plotted the distance (\ref{eq:dLminTem})
as a function of the frequency difference between the template and the signal
expressed in frequency bins. For the case of perfectly matched filter
($\omega_1 = 0$) the distance would have a minimum equal to $0$ for
$\Delta\omega = 0$. For non-zero value of $\omega_1$ the minimum distance is larger
than zero and it occurs for a certain frequency of the filter biased with respect
to true frequency of the signal.

\section{Conclusion}
We have introduced the $L_1$-norm distance $d_L(p_1,p_2)$ between probability density
functions $p_1$ and $p_2$. The $L_1$-norm provides a notion of distance between probability
distributions endowed with a clear probabilistic interpretation, as discussed in Section 3.
The $L_1$-norm distance can be a useful tool in gravitational wave
data analysis for studying the problem of signal resolution, template placements,
and design of search templates. In a future paper we shall study these problems
for realistic gravitational wave signals from supernovae, inspiralling binaries,
rotating neutron stars and white dwarf binaries.

\section*{Acknowledgments}
This work was supported through grant 1~P03B~029~27 of Polish
Ministry of Science and Informatics.

\end{document}